\documentclass[seceq,preprint]{ptptex}

\usepackage{wrapft}

\usepackage{braket,color}
\def\hsymbl#1{\smash{\hbox{\huge$#1$}}}
\def\hsymbu#1{\smash{\lower1.7ex\hbox{\huge$#1$}}}
\newcommand{\bbm}{\begin{bmatrix}}
\newcommand{\ebm}{\end{bmatrix}}
\newcommand{\e}{{\hspace{.1em}{\mathrm e}}}
\newcommand{\afn}[1]{\mathrm{afn} \hspace{-.1em} \left( #1 \right)}
\newcommand{\mrm}{\mathrm}
\newcommand{\tQ}{\tilde{Q}}

\preprintnumber[4cm]{
UT-Komaba/10-10}

\title{
Gauge fixing of open superstring field theory\\[.5ex] in the Berkovits non-polynomial formulation%
}

\author{
Shingo \textsc{TORII}\footnote{
This talk is based on the work in collaboration with Nathan~Berkovits, Michael~Kroyter, Yuji~Okawa, Martin~Schnabl and
Barton~Zwiebach.}%
}

\inst{
Institute of Physics, University of Tokyo,\\
Komaba, Meguro-ku, Tokyo 153-8902, Japan\\
storii@hep1.c.u-tokyo.ac.jp
}

\abst{
We consider gauge fixing of open superstring field theory formulated by Berkovits, concentrating on the Neveu-Schwarz sector.
In the free theory, we perform gauge fixing completely, which requires infinitely many ghosts and 
antighosts carrying various world-sheet ghost numbers and picture numbers.
In the interacting theory, we have determined the form of interactions cubic in fields and antifields
in the Batalin-Vilkovisky formalism.
}

\begin{document}

\maketitle
\section{Introduction}
In string perturbation theory, closed strings appear in loop diagrams of open strings.
However, in the framework of string field theory (SFT), it is non-trivial whether and how closed strings can be described in terms
of open string fields.
Therefore, we would like to know if open SFT can be consistently quantized without additional degrees of 
freedom, such as closed string fields.

In bosonic SFT,\cite{Witten} quantization was discussed\cite{Thorn} by using the Batalin-Vilkovisky (BV) formalism,\cite{BV}
but there was a difficulty caused by divergences from tadpole diagrams.
Moreover, the theory also has difficulty with tachyons: quantization can be considered only in a formal way.
By contrast, in superstring field theory (SSFT), we expect that these difficulties will be absent.
Recent analytic methods in classical bosonic SFT\cite{Witten} which developed mainly 
after the construction of the Schnabl solution\cite{Schnabl}
have been applied to SSFT,
and the work \citen{KZ} by Kiermaier and Zwiebach opened up a vista of extending the methods to at least the one-loop level.
Now we are at the stage for considering quantization of open SSFT seriously.

Various SSFTs have been proposed since the work by Witten.\cite{WittenSuper} 
One approach to formulating an SSFT is using the picture-changing operators\cite{FMS} (PCOs).
At the conference, talks on gauge fixing of the SSFTs of this type, 
\citen{modified} and \citen{Kroyter}, were presented by M.~Murata and M.~Kroyter, respectively.\cite{MK}
Another approach, which is achieved by Berkovits\cite{Berkovits-NS}, is to construct a theory 
in the large Hilbert space without using any PCOs. 
As a step toward quantizing this SSFT, we will first gauge-fix the theory by the BV formalism:
we solve the \textit{master equation}, which is a sort of Ward-Takahashi identity,
and then impose gauge-fixing conditions on the solution.
In the following sections, 
concentrating on the Neveu-Schwarz (NS) sector, we determine gauge-fixing conditions
and solve the equation up to cubic order in fields and antifields.
\section{Gauge Fixing of the Free Theory}
\label{gfft}
The free NS-sector action and its gauge symmetry of the SSFT \citen{Berkovits-NS} are given 
by\footnote{We have appended the subscript ``$0$'' on the action and the gauge variation for later convenience.} 
\begin{subequations}
\begin{align}
S^\mrm{free}_0 &= -\frac{1}{2} \int \Phi_{(0,0)} (Q\eta_0\Phi_{(0,0)}) 
= -\frac{1}{2}\braket{\Phi_{(0,0)}|Q\eta_0 |\Phi_{(0,0)}} \,,
\label{free action} \\
\delta_0 \Phi_{(0,0)} &= Q\epsilon_{(-1,0)} +\eta_0 \epsilon_{(-1,1)}\,.
\label{gdof}
\end{align}
\label{free}
\end{subequations}
Here $\Phi_{(0,0)}$ is a Grassmann-even NS-sector string field,
$Q$ is the BRST operator in the first-quantized theory, and $\eta_0$
is the zero mode of $\eta$, which appears in the bosonization of the superconformal ghosts:
$\beta \cong \e^{-\phi}\partial \xi,\ \gamma \cong \eta \e^\phi$.\cite{FMS}
The integration and the multiplication of string fields are given by Witten's sewing and gluing prescription.\cite{Witten}
The bracket $\braket{\ \ |\ \ }$ is the Belavin-Polyakov-Zamolodchikov (BPZ) inner product:\cite{BPZ}
$\ket{\Phi_{(0,0)}}$ is the state corresponding to the string field $\Phi_{(0,0)}$ and $\bra{\Phi_{(0,0)}}$ is its BPZ conjugate.
In Eq.~(\ref{free}) and in the sequel, the subscript $(g,p)$ on a string field indicates its world-sheet ghost number $g$
and picture $p$.\footnote{The quantum number $(g,p)$ of $\xi$ and that of $\eta$ are
$(-1,1)$ and $(1,-1)$, respectively.} 
Since the theory is formulated in the large Hilbert space, an integral of string fields vanishes unless
the integrand carries $(g,p)=(2,-1)$.

In the free theory, we can easily perform gauge fixing, using the Faddeev-Popov (FP) formalism.
First, we eliminate the gauge symmetries associated with $Q$ and $\eta_0$ by the conditions
$b_0 \ket{\Phi_{(0,0)}}=0$ and $\xi_0 \ket{\Phi_{(0,0)}}=0$, respectively.
The resultant FP action is
\begin{align}
S^\mrm{free}_1 
= \Bigl( \bra{B_{(3,-1)}} b_0 + \bra{B_{(3,-2)}} \xi_0 \Bigr) \Bigl( Q\ket{\Phi_{(-1,0)}} + \eta_0 \ket{\Phi_{(-1,1)}} \Bigr),
\label{S_1 free}
\end{align}
where $\Phi_{(-1,0)}$ and $\Phi_{(-1,1)}$ are ghosts, whereas $B_{(3,-1)}$ and $B_{(3,-2)}$ are antighosts.
If we use the redefined antighost $\Phi_{(2,-1)}$,
\begin{equation}
\bra{\Phi_{(2,-1)}} := \bra{B_{(3,-1)}} b_0 + \bra{B_{(3,-2)}} \xi_0 \,,
\end{equation}
we obtain the gauge-fixed action of the form
\begin{equation}
S^\mrm{free}_0 + S^\mrm{free}_1 = -\frac{1}{2}\braket{\Phi_{(0,0)}|Q\eta_0 |\Phi_{(0,0)}}
+ \bra{\Phi_{(2,-1)}} \Bigl( Q\ket{\Phi_{(-1,0)}} + \eta_0 \ket{\Phi_{(-1,1)}} \Bigr)
\label{free S_0+S_1}
\end{equation}
with
\begin{equation}
\bbm
b_0 \\
\xi_0 
\ebm
\Phi_{(0,0)} = 0\,,\quad b_0 \xi_0 \Phi_{(2,-1)}=0\,.
\label{constraints}
\end{equation}
Decomposing the string fields with respect to the zero modes $c_0$ and $\xi_0$ helps one
understand that the condition (\ref{constraints}) really eliminates the gauge degree of freedom (\ref{gdof}).
However, gauge fixing of the theory has not been completed yet.
Since $Q$ and $\eta_0$ satisfy the relation
\begin{equation}
Q^2=\eta_0^2=\{ Q,\eta_0\} =0\,,
\label{Qeta}
\end{equation}
the action $S^\mrm{free}_0 + S^\mrm{free}_1$ is invariant under the gauge transformation of the ghosts
\begin{subequations}
\begin{align}
\delta_1 \Phi_{(-1,0)} &= Q\epsilon_{(-2,0)}+ \eta_0\epsilon_{(-2,1)}\,,\\
\delta_1 \Phi_{(-1,1)} &= Q\epsilon_{(-2,1)}+ \eta_0\epsilon_{(-2,2)}\,.
\end{align}
\label{ghost gauge transf}
\end{subequations}
To remove this symmetry, we use the gauge-fixing condition of the form
\begin{equation}
\bbm
b_0 & 0\\
\xi_0 & b_0 \\
0 & \xi_0
\ebm
\bbm
\Phi_{(-1,0)} \\
\Phi_{(-1,1)}
\ebm
=0\,,
\end{equation}
and introduce the ghosts for ghosts $\Phi_{(-2,0)}$, $\Phi_{(-2,1)}$ and $\Phi_{(-2,2)}$.
The resultant FP action is
\begin{align}
S^\mrm{free}_2 &= \bra{B_{(4,-1)}} b_0 \Bigl( Q\ket{\Phi_{(-2,0)}} + \eta_0 \ket{\Phi_{(-2,1)}} \Bigr) \nonumber \\
&+ \bra{B_{(4,-2)}} \Bigl[ \xi_0 \Bigl( Q\ket{\Phi_{(-2,0)}} + \eta_0 \ket{\Phi_{(-2,1)}} \Bigr)
+b_0 \Bigl( Q\ket{\Phi_{(-2,1)}} + \eta_0 \ket{\Phi_{(-2,2)}} \Bigr) \Bigr] \nonumber \\
&+ \bra{B_{(4,-3)}} \xi_0 \Bigl( Q\ket{\Phi_{(-2,1)}} + \eta_0 \ket{\Phi_{(-2,2)}} \Bigr) \nonumber \\
&= \bra{\Phi_{(3,-1)}} \Bigl( Q\ket{\Phi_{(-2,0)}} + \eta_0 \ket{\Phi_{(-2,1)}} \Bigr)
+ \bra{\Phi_{(3,-2)}} \Bigl( Q\ket{\Phi_{(-2,1)}} + \eta_0 \ket{\Phi_{(-2,2)}} \Bigr)\,,
\end{align}
where
\begin{align}
&\bra{\Phi_{(3,-1)}} := \bra{B_{(4,-1)}} b_0 + \bra{B_{(4,-2)}} \xi_0\,,\quad 
\bra{\Phi_{(3,-2)}} := \bra{B_{(4,-2)}} b_0 + \bra{B_{(4,-3)}} \xi_0\,,\\
&\bbm
b_0 & \xi_0
\ebm
\bbm
\Phi_{(3,-1)} \\
\Phi_{(3,-2)}
\ebm
=0\,,\quad 
b_0 \xi_0 \ket{\Phi_{(3,-1)}} = b_0 \xi_0 \ket{\Phi_{(3,-2)}} =0\,.
\end{align}
The gauge-fixed action so far is
\begin{align}
\sum_{k=0}^2 S^\mrm{free}_k &= -\frac{1}{2}\braket{\Phi_{(0,0)}|Q\eta_0 |\Phi_{(0,0)}}
+ \bra{\Phi_{(2,-1)}} \bigl( Q\ket{\Phi_{(-1,0)}} + \eta_0 \ket{\Phi_{(-1,1)}} \bigr) \nonumber \\
&+ \bra{\Phi_{(3,-1)}} \bigl( Q\ket{\Phi_{(-2,0)}} + \eta_0 \ket{\Phi_{(-2,1)}} \bigr)
+ \bra{\Phi_{(3,-2)}} \bigl( Q\ket{\Phi_{(-2,1)}} + \eta_0 \ket{\Phi_{(-2,2)}} \bigr)
\end{align}
with the constraints.

In this way, we can continue gauge fixing step by step and obtain the completely-gauge-fixed action
\begin{align}
S^\mrm{free} &= -\frac{1}{2} \int \Phi_{(0,0)} (Q\eta_0\Phi_{(0,0)})
+ \sum^\infty_{n=1} \sum^{n-1}_{m=0} \int \Phi_{(n+1,-m-1)} \bigl( Q\Phi_{(-n,m)} + \eta_0 \Phi_{(-n,m+1)} \bigr)
\end{align}
with
\begin{subequations}
\begin{align}
&\underbrace{
\begin{bmatrix}
b_0 & & \hsymbu{0} \\
\xi_0 & \ddots & \\
 & \ddots & b_0 \\
\hsymbl{0} & & \xi_0 \\
\end{bmatrix}
}_{n+1}
\begin{bmatrix}
\Phi_{\left( -n ,\, 0 \right)} \\
\vdots \\
\vdots \\
\Phi_{\left( -n ,\, n \right)} 
\end{bmatrix}
=0
\quad \left( n \geq 0 \right),\\
&\underbrace{
\begin{bmatrix}
b_0 & \xi_0 & & \hsymbu{0} \\
  & \ddots & \ddots & \\
\hsymbl{0} & & b_0 & \xi_0 \\
\end{bmatrix}
}_{n-1}
\begin{bmatrix}
\Phi_{\left( n ,\, -1 \right)} \\
\vdots \\
\Phi_{\left( n ,\, -\left(n-1\right) \right)} 
\end{bmatrix}
=0
\quad \left( n \geq 3 \right),\\
&b_0 \xi_0 \Phi_{(n,-m)} =0 \quad \left( 1\leq m \leq n-1 \right).
\end{align}
\label{gf cond}
\end{subequations}
\begin{wrapfigure}{l}{\halftext}
\unitlength 0.1in
\begin{picture}( 22.8000, 14.6500)( 11.2000,-26.9500)
%
{\color[named]{Black}{%
\special{pn 20}%
\special{pa 2000 1996}%
\special{pa 3400 1996}%
\special{fp}%
\special{sh 1}%
\special{pa 3400 1996}%
\special{pa 3334 1976}%
\special{pa 3348 1996}%
\special{pa 3334 2016}%
\special{pa 3400 1996}%
\special{fp}%
\special{pa 2600 2596}%
\special{pa 2600 1396}%
\special{fp}%
\special{sh 1}%
\special{pa 2600 1396}%
\special{pa 2580 1462}%
\special{pa 2600 1448}%
\special{pa 2620 1462}%
\special{pa 2600 1396}%
\special{fp}%
}}%
%
{\color[named]{Black}{%
\special{pn 13}%
\special{pa 2600 1996}%
\special{pa 2000 1396}%
\special{fp}%
}}%
%
{\color[named]{Black}{%
\special{pn 13}%
\special{pa 2800 2096}%
\special{pa 3400 2696}%
\special{fp}%
}}%
%
{\color[named]{Black}{%
\special{pn 8}%
\special{pa 2290 1706}%
\special{pa 2010 1986}%
\special{fp}%
\special{pa 2260 1676}%
\special{pa 2000 1936}%
\special{fp}%
\special{pa 2230 1646}%
\special{pa 2000 1876}%
\special{fp}%
\special{pa 2200 1616}%
\special{pa 2000 1816}%
\special{fp}%
\special{pa 2170 1586}%
\special{pa 2000 1756}%
\special{fp}%
\special{pa 2140 1556}%
\special{pa 2000 1696}%
\special{fp}%
\special{pa 2110 1526}%
\special{pa 2000 1636}%
\special{fp}%
\special{pa 2080 1496}%
\special{pa 2000 1576}%
\special{fp}%
\special{pa 2050 1466}%
\special{pa 2000 1516}%
\special{fp}%
\special{pa 2020 1436}%
\special{pa 2000 1456}%
\special{fp}%
\special{pa 2320 1736}%
\special{pa 2070 1986}%
\special{fp}%
\special{pa 2350 1766}%
\special{pa 2130 1986}%
\special{fp}%
\special{pa 2380 1796}%
\special{pa 2190 1986}%
\special{fp}%
\special{pa 2410 1826}%
\special{pa 2250 1986}%
\special{fp}%
\special{pa 2440 1856}%
\special{pa 2310 1986}%
\special{fp}%
\special{pa 2470 1886}%
\special{pa 2370 1986}%
\special{fp}%
\special{pa 2500 1916}%
\special{pa 2430 1986}%
\special{fp}%
\special{pa 2530 1946}%
\special{pa 2490 1986}%
\special{fp}%
}}%
%
{\color[named]{Black}{%
\special{pn 4}%
\special{pa 3390 2126}%
\special{pa 3120 2396}%
\special{fp}%
\special{pa 3390 2186}%
\special{pa 3150 2426}%
\special{fp}%
\special{pa 3390 2246}%
\special{pa 3180 2456}%
\special{fp}%
\special{pa 3390 2306}%
\special{pa 3210 2486}%
\special{fp}%
\special{pa 3390 2366}%
\special{pa 3240 2516}%
\special{fp}%
\special{pa 3390 2426}%
\special{pa 3270 2546}%
\special{fp}%
\special{pa 3390 2486}%
\special{pa 3300 2576}%
\special{fp}%
\special{pa 3390 2546}%
\special{pa 3330 2606}%
\special{fp}%
\special{pa 3390 2606}%
\special{pa 3360 2636}%
\special{fp}%
\special{pa 3360 2096}%
\special{pa 3090 2366}%
\special{fp}%
\special{pa 3300 2096}%
\special{pa 3060 2336}%
\special{fp}%
\special{pa 3240 2096}%
\special{pa 3030 2306}%
\special{fp}%
\special{pa 3180 2096}%
\special{pa 3000 2276}%
\special{fp}%
\special{pa 3120 2096}%
\special{pa 2970 2246}%
\special{fp}%
\special{pa 3060 2096}%
\special{pa 2940 2216}%
\special{fp}%
\special{pa 3000 2096}%
\special{pa 2910 2186}%
\special{fp}%
\special{pa 2940 2096}%
\special{pa 2880 2156}%
\special{fp}%
\special{pa 2880 2096}%
\special{pa 2850 2126}%
\special{fp}%
}}%
%
{\color[named]{Black}{%
\special{pn 13}%
\special{pa 3400 2096}%
\special{pa 2800 2096}%
\special{fp}%
}}%
\put(20.0000,-13.9500){\makebox(0,0)[rb]{{\small $g+p=0$}}}%
\put(34.0000,-26.9500){\makebox(0,0)[lt]{{\small $g+p=1$}}}%
\put(28.0000,-18.9500){\makebox(0,0){{\small $2$}}}%
%
{\color[named]{Black}{%
\special{pn 13}%
\special{pa 2800 1996}%
\special{pa 2800 2096}%
\special{dt 0.045}%
\special{pa 2800 2096}%
\special{pa 2600 2096}%
\special{dt 0.045}%
}}%
\put(24.6000,-20.9000){\makebox(0,0){{\small $-1$}}}%
\put(26.0000,-12.9500){\makebox(0,0){{\small $p$}}}%
\put(35.0000,-19.9500){\makebox(0,0){{\small $g$}}}%
\end{picture}%
\\[-1.5ex]
\caption{Distribution of allowed $(g,p)$.}
\label{figure}
\end{wrapfigure}
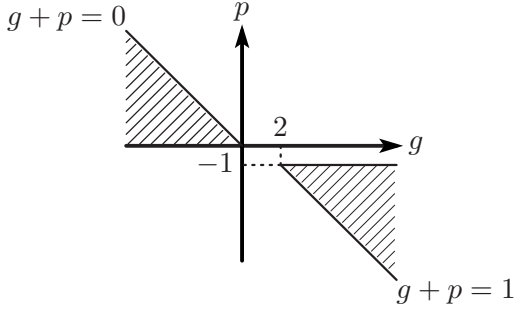
Here $\Phi_{(g,p)}$ with $g\leq -1,\ 0\leq p \leq -g$ are ghosts and those with $-(g-1)\leq p \leq -1$ are antighosts.
Note that a string field $\Phi_{(g,p)}$ is admissible only when the lattice point $(g,p)$ belongs to the region
shown in Fig.~\ref{figure}: the original field $\Phi_{(0,0)}$ and the ghosts live in the left region, whereas the antighosts live in
the right region in the figure.
\section{Interacting Theory}
The full NS-sector action in ref.~\citen{Berkovits-NS} and its gauge symmetry are of the
Wess-Zumino-Witten type:
\begin{subequations}
\begin{align}
S_0 &= \frac{1}{2} \int \biggl( G^{-1} ( QG ) G^{-1} ( \eta_0 G)
- \int_{0}^{1} \mathrm{d}t \left({\hat{G}}^{-1} {\partial}_t \hat{G} \right) 
\left\{ {\hat{G}}^{-1} ( Q \hat{G} ) , {\hat{G}}^{-1} ( \eta_0 \hat{G} ) \right\} \biggr) ,
\label{full action} \\
\delta_0 G &= \left( Q\epsilon_{(-1,0)}\right) G + G\left( \eta_0 \epsilon_{(-1,1)}\right),
\label{G gauge} \\ 
G &= \exp(\Phi_{(0,0)}),\ \hat{G} = \exp(t\Phi_{(0,0)})\,.
\nonumber
\end{align}
\end{subequations}
Note that the integral with respect to the variable $t$ is an ordinary integral. 
In the interacting theory, it is difficult to complete gauge fixing
through the FP procedure. This is because the FP action 
has gauge invariance which requires the equation of motion.
To see this, it is convenient to set 
\begin{equation}
\epsilon_{(-1,0)} = \e^{\phi} \tilde{\epsilon}_{(-1,0)} \e^{-\phi}\,,\quad 
\phi := \Phi_{(0,0)}\,,
\end{equation}
and redefine $\tilde{\epsilon}_{(-1,0)}$  as $\epsilon_{(-1,0)}$. This leads to
\begin{align}
\delta_0 \e^{\phi}
= \e^{\phi} \bigl( \tQ \epsilon_{(-1,0)} + \eta_0 \epsilon_{(-1,1)} \bigr),
\label{s3}
\end{align}
where
\begin{align}
\tQ X 
:= \e^{-\phi} Q \bigl( \e^{\phi} X \e^{-\phi} \bigr) \e^{\phi}
= QX + \Bigl[ \e^{-\phi}\bigl( Q\e^{-\phi}\bigr), X\Bigr] .
\end{align}
Here and in the sequel, the bracket $[\ ,\ ]$ means the graded commutator.
In the free case, the action (\ref{free S_0+S_1}) is exactly invariant under the transformation
(\ref{ghost gauge transf}) because Eq.~(\ref{Qeta}) holds.
In the interacting case, however, the relation corresponding to Eq.~(\ref{Qeta}) is
\begin{equation}
\tQ^2 = \eta_0^2 = 0\,,\quad
\{ \tQ, \eta_0 \} X = \left[ \frac{\delta S_0}{\delta \e^{\phi}} \e^{\phi}, X\right],
\end{equation}
so that $\{ \tQ, \eta_0 \}$ vanishes only when we use the equation of motion.
This makes it difficult to carry out the FP procedure (or the BRST procedure).
However, there is a powerful formalism to deal with such a complicated gauge system systematically.
It is the BV formalism,\cite{BV} and we use it to consider gauge fixing of the interacting theory.
\section{The BV Formalism and Antifield Number Expansion}
The BV formalism is an extension of the BRST formalism.
In general, gauge fixing in the BV formalism is performed as follows.\footnote{We use 
the notation where the appearance of discrete index also indicates the presence of a spacetime variable.}
First, one prepares ghosts, ghosts for ghosts, and so on as well as the original fields $\phi^i$ in a given action $S_0[\phi]$.
Second, for each field $\Phi^A \in \{\phi^i,\ \text{ghosts, ghosts for ghosts, ...}\}$, 
one introduces an additional field called an \textit{antifield} $\Phi^\ast_A$.
Third, starting with the action $S_0[\phi]$, one constructs the solution $S[\Phi,\Phi^\ast]$ to the \textit{master equation}
\begin{equation}
\sum_A \frac{\partial_r S}{\partial \Phi^A} \frac{\partial_l S}{\partial \Phi^\ast_A} = 0
\end{equation}
under the boundary condition
\begin{equation}
S[\Phi,\Phi^\ast]|_{\Phi^\ast =0} = S_0[\phi] \,.
\end{equation}
Here $\partial_l$ and $\partial_r$ are the left and the right derivative, respectively.
Fourth, one eliminates the gauge symmetry of $S$, which originates from the symmetries of $S_0$ and ghost actions.
This is achieved by imposing on the antifields conditions of the form
\begin{equation}
\Phi^\ast_A = \frac{\partial \Psi[\Phi]}{\partial \Phi^A} \,,
\label{Cond}
\end{equation}
where $\Psi[\Phi]$ is a functional of $\Phi^A$'s.
Finally, one obtains the completely-gauge-fixed action $S[\Phi,{\partial \Psi}/{\partial \Phi} ]$.
In the BV formalism, unlike in the BRST formalism, one can remove all the gauge degrees of freedom at once
including those associated with ghosts, by the condition (\ref{Cond}).
Moreover, the BRST invariance of the gauge-fixed action is essentially equivalent to the master equation.

Let us apply this formalism to the SSFT.
Note that the action $S^\mrm{free}$ (with no constraints) obtained in the free theory satisfies the master equation of the form
\begin{equation}
\sum^\infty_{n=0} \sum^n_{m=0} \int \frac{\delta_r S^\mrm{free}}{\delta \Phi_{(-n,m)}}\ \frac{\delta_l S^\mrm{free}}{\delta \Phi_{(n+2,-m-1)}} =0\,.
\label{free master}
\end{equation}
Actually, we can choose gauge-fixing conditions imposed on antifields 
such that the antifield of $\Phi_{(0,0)}$ and those of the ghosts are identified with the antighosts\footnote{Strictly speaking, 
under the identification (\ref{identify}),
Eq.~(\ref{free master}) corresponds to the master equation for 
the \textit{minimal set} of fields, which consists of the original fields and ghosts.
Not until the \textit{non-minimal} set is introduced and Lagrange multiplier fields are integrated out,
is the gauge-fixing condition (\ref{gf cond}) imposed. For detail, see ref.~\citen{GPS}, for example.}
\begin{equation}
\Phi^\ast_{(g,p)} \cong \Phi_{(2-g,-1-p)}\,,\quad ( 0\leq p \leq -g)\,.
\label{identify}
\end{equation}
Therefore, in what follows, we do not have to distinguish them.

In the interacting theory, it is not so easy to construct the solution to the master equation as in the free theory.
However, it is known that one can solve the equation step by step, using the expansion in 
\textit{antifield number}, which is defined in the present case as
\begin{equation}
\afn{\Phi_{(g,p)}} := \left\{
\begin{aligned}
&0 \quad \left( 0\leq p \leq -g \right) \\
&g-1 \quad \left( -(g-1)\leq p \leq -1 \right)
\end{aligned}
\right. \,.
\end{equation}
For example, the antifield numbers of $S^\mrm{free}_0$, $S^\mrm{free}_1$ and $S^\mrm{free}_2$ 
in section \ref{gfft} are zero, one and two, respectively.
We expand the solution $S$ in antifield number,
\begin{equation}
S = \sum^\infty_{k=0} S_k,\quad \afn{S_k} =k\,.
\end{equation}
Our strategy for solving the master equation in the interacting theory is as follows.
\begin{itemize}
\item First, we solve the master equation to some antifield number.
\item Second, from the result of the first step, we infer the complete solution and confirm its validity.
\end{itemize}
In the bosonic SFT \citen{Witten}, the second step is easy: the complete solution can be obtained
only by removing the world-sheet ghost number constraint imposed on the string field.
In the present SSFT, however, this is not the case.
As we can see from the result in the free theory,
the original action $S^\mrm{free}_0$ contains the term involving the \textit{product} of $Q$ and $\eta_0$, whereas
the FP terms do not: they consist of the \textit{sum} of $Q$-terms and $\eta_0$-terms.
Therefore there seems to be no chance that the solution will take the same form as the original action. 

Starting with Eqs.~(\ref{full action}) and (\ref{s3}), we have
\begin{subequations}
\begin{align}
S_1 &=
\int \Phi_{(2,-1)} \e^\phi \bigl( \tilde{Q} \Phi_{(-1,0)} + \eta_0 \Phi_{(-1,1)} \bigr),\\ 
S_2 &=
\int \biggl[ \Phi_{(2,-1)} \e^\phi \Phi_{(2,-1)} \e^\phi \Phi_{(-2,1)}
\nonumber \\
&\quad + 
\Phi_{(3,-1)} \Bigl( \tilde{Q} \Phi_{(-2,0)} + \eta_0 \Phi_{(-2,1)} +
\Phi_{(-1,0)} \tilde{Q} \Phi_{(-1,0)} + \bigl[ \Phi_{(-1,0)}, \eta_0 \Phi_{(-1,1)} \bigr]
\Bigr)
\nonumber \\
&\quad + \Phi_{(3,-2)} \Bigl( \tilde{Q} \Phi_{(-2,1)} + \eta_0 \Phi_{(-2,2)} +
\Phi_{(-1,1)} \eta_0 \Phi_{(-1,1)}
\Bigr) \biggr] .
\end{align}
\end{subequations}
We have calculated $S_3$ and $S_4$ as well, but have not been able to infer a complete solution so far. 
For cubic interactions, however, a complete form has been obtained, as we will show
in the next section.
\section{Cubic Interactions}
It is known that the solution to the classical master equation is unique only up to \textit{canonical transformations},
which correspond to a subset of the degrees of freedom of field redefinition.
We expect that an appropriate field redefinition will help us infer a solution.
In order to find such a field redefinition, we use the $\mathbb{Z}_2$-transformation property of the original action
(\ref{full action}): the action $S_0$ becomes $-S_0$ under the $\mathbb{Z}_2$-transformation
\begin{equation}
\left(\Phi_{(0,0)},Q,\eta_0 \right) \rightarrow \left(-\Phi_{(0,0)},\eta_0 ,Q\right).
\end{equation}
Let us extend this transformation to all the ghosts and antighosts obtained in the free case. 
We readily find that $S^\mrm{free}$ is transformed into $-S^\mrm{free}$ under
\begin{equation}
\left(\Phi_{(g,p)}, \Phi_{(2-g,-1-p)}, Q, \eta_0 \right) \rightarrow
\left(-\Phi_{(g,-g-p)}, +\Phi_{(2-g,g+p-1)}, \eta_0 ,Q\right) \quad (0\leq p \leq -g) \,.
\label{exZ2}
\end{equation}
Using the antifield number expansion, we have constructed the solution up to cubic order in fields and antifields
which respects this extended $\mathbb{Z}_2$-transformation property.
If we write the coupling constant $g$ explicitly, the result takes the form
\begin{align}
S &= S^\mrm{free} + g\, S^\mrm{cubic} +O(g^2)\,,\\
S^\mrm{cubic}
&= \int \biggl[ - \frac{1}{6}\phi\bigl[Q\phi,\eta\phi\bigr]
+\Phi^\ast \Bigl( \frac{1}{2} \bigl[Q\phi,\Phi \bigr]
-\frac{1}{2}\bigl[\eta\phi, \Phi \bigr] \Bigr)
+ \Phi^\ast \Phi^\ast \Phi
\nonumber \\
&+\frac{1}{2} \sum^\infty_{n=1} \sum^\infty_{m=n+2} \sum^{m-n-1}_{k=1}
\Bigl( \bigl[\Phi_{(m,-k)},\Phi_{(1+n-m,-1+k)}\bigr] \eta \Phi_{(-n,1)} \nonumber \\
&\qquad \qquad \qquad \qquad -\bigl[\Phi_{(m,-m+k)},\Phi_{(1+n-m,-n+m-k)}\bigr] Q\Phi_{(-n,n-1)} \Bigr)
\biggr] .
\label{S^(3)}
\end{align}
Here $\Phi$ and $\Phi^\ast$ are the sum of all the ghosts and the sum of all the antighosts, respectively,
\begin{equation}
\Phi := \sum^\infty_{n=1} \sum^n_{m=0} \Phi_{(-n,m)}\,,\quad
\Phi^\ast := \sum^\infty_{n=2} \sum^{n-1}_{m=1} \Phi_{(n,-m)}\,.
\end{equation}
We can easily confirm that this action $S$ is transformed into $-S$ under the extended $\mathbb{Z}_2$-transformation
(\ref{exZ2}).
Moreover, $S$ is really a solution to the master equation in the following sense:
\begin{equation}
\sum^\infty_{n=0} \sum^n_{m=0} \int \frac{\delta_r S}{\delta \Phi_{(-n,m)}}\ \frac{\delta_l S}{\delta \Phi_{(n+2,-m-1)}}
= O(g^2) .
\end{equation}
If we impose on $S$ the gauge-fixing conditions given in section \ref{gfft},
we obtain the gauge-fixed action at this order. 

We have now determined the cubic interactions in the NS sector completely.
If we have those in the Ramond sector as well, we can extend the analysis \citen{EST}
by Ellwood, Shelton and Taylor to the SSFT.
Then it would be exciting to see if the gauge invariance is preserved or anomalous for one-point functions at one loop,
which contain closed strings.

\section*{Acknowledgements}
This work was supported in part by Research Fellowships of the Japan Society for the Promotion of 
Science for Young Scientists.

%

\end{document}